# Experimental demonstration of the slow group velocity of light in two-dimensional coupled photonic crystal microcavity arrays

Hatice Altug[*] and Jelena Vučković[†]

*Edward L. Ginzton Laboratory, Stanford University, Stanford, CA 94305-4088*

**Abstract**

We recently proposed two-dimensional coupled photonic crystal microcavity arrays as a route to achieve a slow-group velocity of light (flat band) in all crystal directions. In this paper we present the first experimental demonstration of such structures with a measured group velocity below $0.008c$ and discuss the feasibility of applications such as low-threshold photonic crystal lasers with increased output powers, optical delay components and sensors.

[*] Also at the Department of Applied Physics, Stanford University, Stanford, CA 94305, altug@stanford.edu
[†] Also at the department of Electrical Engineering, Stanford University, Stanford, CA 94305, jela@stanford.edu, http://www.stanford.edu/group/nqp

One-dimensional coupled resonator optical waveguides (1D CROWs) had been proposed by Yariv et al. in 1999 [1] and subsequently demonstrated experimentally in microwave and optical domains [2-4]. 1D CROWs attracted a significant attention because of their ability to provide a very small group velocity of light in the direction in which the cavities are coupled. As in the case of photonic crystal (PhC) waveguides, the coupling into such structures is difficult, since the input beam needs to be aligned in one particular direction and coupled efficiently into a cross-sectional area that is typically much smaller than 1μm$^2$. On the other hand, PhC microcavity lasers have been studied by many groups because of their small size, low threshold and possibility of high direct modulation rate [5-8]; unfortunately, their output powers are very low for most applications.

We have recently proposed two-dimensional coupled photonic crystal resonator arrays (2D CPCRAs), which enable a small group velocity of light for all wavevectors and in all crystal directions with a very low group velocity dispersion [9]. Our design overcomes the noted disadvantages of 1D CROWs: since a small group velocity occurs in all propagation directions, the coupling into the structure is thus not as difficult. Moreover, when the structure is combined with an active material and employed as a laser, a much higher output power can be achieved than in PhC microcavity lasers, while preserving a low threshold and a high-direct modulation rate. A high density of optical states in such structures can also be used to study nonlinear optics at low thresholds, or to construct compact polarizing optical elements and sensors. A 2D CPCRA is formed by tiling microcavities in two dimensions, as shown in Fig. 1b. When combined into a two-dimensional network, defect modes of individual cavities form coupled defect bands located inside the photonic bandgap [9]. In particular, the coupled arrays in the square lattice photonic crystal exhibit three coupled TE-like bands: monopole, dipole, and quadrupole [9]. The coupled monopole band is close to the air band edge and is not flat; this band is also outside the wavelength range probed in our experiments and is not studied in this paper. z-components of the magnetic field at the center of the PhC slab for the coupled dipole and quadrupole modes are shown in the insets of Figs. 2 and 3, respectively. The coupled dipole bands originate from the doubly degenerate dipole mode; two sub-bands corresponding to the coupled x- and the y-dipoles split away from the Γ point in the ΓX direction, while one degenerate diagonal dipole band is observed in the ΓM direction. Because of the preferential radiation directions for the dipole mode, the coupled dipole bands exhibit different group velocities in different crystal directions. Conversely, the coupled quadrupole band exhibits a small group velocity in all crystal directions, as it originates from the non-degenerate quadrupole mode with a large in-plane Q-factor and with a four-fold symmetric radiation pattern. In this article, we present experimental studies of both dipole and quadrupole bands.

The designed 2D CPCRAs were fabricated in silicon on insulator (SOI), as described in our earlier publication [9], and scanning electron microscope (SEM) pictures of the fabricated structures are shown in Fig. 1b. For the experiments presented here, a low-index silicon-dioxide ($SiO_2$) layer underneath the silicon (Si) membranes was not undercut. The presence of $SiO_2$ layer red-shifts resonance frequencies and mixes the TE and TM polarizations, but does not affect the slope of the bands significantly, as confirmed by 3D Finite-Difference Time-Domain (FDTD) simulations. The parameters of the fabricated structures are: the PhC periodicity $a$ = 490 nm, the hole radius $r$ = 190nm, the slab thickness $d$ = 275 nm, and the CPCRA unit cell dimensions $A \times A = 3a \times 3a$ (i.e., two PhC layers are inserted between tiled microcavities). The experimental setup used in the measurement of the dispersion diagram is shown in Fig. 1a. The structure is excited by a tunable laser source with a scanning range of 1460nm-1580nm. The beam is slightly focused on the sample by a very low numerical aperture (NA) objective lens, so that the excitation can be approximated by a plane wave; this is important for the control of the in-plane wave-vector (k-vector). The size of the device is 100μm×100μm and the pump beam has the diameter of 120μm at the sample surface. Polarizing optics is employed to preferentially excite TE-like (even) modes of the structure. We have tested both transmission and reflection spectra of the structure with same and opposite polarization relative to the pump laser. The in-plane k-vector and the choice of the PhC symmetry direction (ΓX or ΓM) are controlled by rotating the structure around the y- and z-axes, respectively. In order to position the pump beam relative to the device, its reflection is imaged; therefore, the maximum tilt angle that can be tested (before losing the image) is determined by the collection optics and is equal to 13°. However, this is sufficient to probe a relevant portion of the studied CPCRA dispersion diagram, since the tilt angle $\theta_X$ required to reach the X point of the 2-layer CPCRA is around 31° as estimated from $k_\parallel = \frac{2\pi}{\lambda}\sin(\theta_X) = \frac{\pi}{3a}$ and $a/\lambda \approx 0.3$. Due to the low NA of the objective lens, the collected reflected signal drops rapidly when the sample tilt around the y-axis increases; the transmitted signal is thus more convenient for measuring the band diagram. The inset of Fig. 2a shows the transmission spectrum of the dipole band at the Γ point (corresponding to the 0° tilt), with the incident beam polarized in the x-direction, thereby preferentially exciting the x-dipole. A strong resonance dip near 1564nm can be observed, indicating the spectral position of the mode. By tracking the position of the dip at different tilt angles, we can extract a full band diagram. It should be emphasized that the linewidth of the spectrum is very small (7nm), even though the signal is transmitted through an array containing 3600 coupled cavities, indicating a high uniformity of the fabricated structures. The 7nm linewidth corresponds to a quality factor (Q) of 220 while for the dipole mode in a single-defect microcavity embedded in a square PhC lattice Q≈300.

The transmission spectra measured at different tilt angles in the ΓX direction with pump polarized in the x- and y-directions are plotted in Figs. 2a and 2b, respectively. The dark blue stripes located between 1545-1565nm in these figures correspond to the positions of the coupled bands: the flat x-dipole band (Fig. 2a) and the non-flat y-dipole band (Fig. 2b). We note two discrepancies relative to the theory [9]: first, the x- and y-dipole bands do not have the same wavelengths at the Γ point, and second, a different set of x- and y-dipole bands is observed when the same set of measurements is performed on the structure rotated by 90° around the z-axis. These two observations indicate that fabricated structures are not symmetric under 90° rotation. This asymmetry can be noticed in SEM pictures of the fabricated structures: structures are slightly astigmated (as a result of imperfect electron beam lithography), with hole diameter in the x-direction larger by roughly 20nm (5%) than in the y-direction (the structure periodicity is approximately equal in both directions). The degeneracy of the dipole modes is also lifted in the ΓM direction, where the diagonal dipole splits into x-like and y-like dipole bands. On the other hand, the quadrupole mode is non-degenerate and the corresponding coupled band does not split into subbands due to structure imperfections. We have studied the structures astigmated by 6.25% numerically by employing FDTD with discretization of 20 units per *a*; results for dipole bands in the ΓX and ΓM directions are plotted in Fig. 3a and Fig. 3b, respectively, together with corresponding experimental results (determined by fitting inverted Lorentzians to transmission dips). A very good match between experiment and theory is observed, and a small discrepancy results mainly from a modest FDTD discretization. The wavelength resolutions in simulation and experiment are 5.5nm and 0.2nm-1nm, respectively, and error bar for simulation is indicated in Fig. 3. We have also measured the reflected signal at opposite polarization relative to the pump laser at the Γ-point, which exhibits a strong peak at the location of the coupled band. Such polarization conversion in CPCRAs can be employed in building compact polarizing optical elements or sensors, as will be discussed in our forthcoming publication.

The coupled quadrupole band has not been observed under plane-wave excitation described above, although theory predicts that it should be in the scanned wavelength range. The reason is that the plane-wave excitation is linearly polarized and uniform in the x-y plane (at or close to the Γ point), so it cannot excite the quadrupole mode whose electric field has an odd symmetry and an equal presence of the $E_x$ and $E_y$ components. In order to approximately locate the position of the quadrupole band, the structure is tilted by approximately 20° around the y-axis and the beam is tightly focused; in this case, the excitation field varies in the plane of the structure and is no longer orthogonal to the mode. However, this approach cannot be used to map the quadrupole band accurately, as the k-vector information is lost due to tight focus. The experimental result is shown in Fig. 3c, and it is very close to the location of the quadrupole band predicted by FDTD simulation. In an active CPCRA device (e.g., a laser containing quantum wells), photons emitted inside the structure can excite modes with arbitrary in-plane symmetry, so the coupling to the quadrupole band would not be a problem.

As can be seen in the simulated and measured band diagrams, the group velocity of the coupled x-dipole band is small for all wavevectors in the ΓX direction. The measured group velocity at the Γ point is below: $v_g|_{max} = \frac{\Delta\omega}{\Delta k} = \frac{\Delta\lambda}{\lambda}\frac{1}{\Delta\theta} \leq 8\times 10^{-3} c$ where $\Delta\theta=\pi/180$ is the tilt step in radians, $\Delta\lambda=0.2$nm is wavelength resolution, $\lambda=1564$nm is the resonance wavelength and $c$ is the speed of light in vacuum. For optical delay applications, a coupled quadrupole band would be even more suitable, since it exhibits a small group velocity over all k-vectors and in all directions [9]. For example, for a symmetric and undercut CPCRA with $A=4a$, we estimate theoretically that the group velocity of the quadrupole band is below $0.0138c$ and $0.04c$ in the whole ΓX and ΓM direction, respectively. These results were obtained by fitting cosine functions (as predicted for the CPCRA band diagram by the tight binding method [1]) between the quadrupole band frequencies at the high symmetry points (Γ, X and M) calculated by the FDTD method.

Finally, we evaluate advantages of a CPCRA laser relative to a PhC microcavity laser. First, its output power would be approximately N times larger, as it would operate as a phase-locked array of N microlasers. Second, its electrical pumping would not be as challenging, since approaches similar to those employed in pumping large area PhC band-edge lasers could be used [11]. Lastly, a pump beam is generally much larger than the cavity size in a PhC microcavity laser, and majority of excitons are created in the mirror regions where they cannot couple to the cavity mode, thereby degrading the spontaneous emission coupling factor and the lasing threshold [12]. However in the CPCRA laser the pump beam overlaps with many cavities and the wasted pump power is reduced.

In conclusion, we have experimentally demonstrated the band diagram of the 2D CPCRAs that we proposed recently [9], and confirmed the existence of flat bands inside the photonic bandgap independent of light propagation direction, with measured group velocities below $0.008c$. We are currently working on applications of these structures as lasers and sensors.

**Acknowledgements:** This work has been supported by the Marco Interconnect Focus Center, and in part by the MURI Center for Photonic Quantum Information Systems (ARO/ARDA program DAAD19-03-1-0199), the Charles Lee Powell Foundation Faculty Award and the Terman Award. The authors would like to thank Mehmet F. Yanik for discussion of the experiment, and Dr. Jim McVittie and Eric Perozziello for discussions of dry etching.

**References:**

[1] A. Yariv, Y. Xu, R. K. Lee, A. Scherer, Optics Lett. **24**, 711 (1999)

[2] M. Bayindir, B. Temelkuran, E. Ozbay, Phys. Rev. Lett. **84**, 2140 (2000)

[3] S. Olivier, C. Smith, M. Rattier, H. Benisty, C. Weisbuch, T. Krauss, R. Houdre, U. Oesterle, Optics Lett. **26**, 1019 (2001)

[4] T. D. Happ, M Kamp, A. Forchel, J. Gentner, L. Goldstein, Appl. Phys. Lett. **82**, 4 (2003)

[5] O. Painter, R.K. Lee, A. Scherer, A. Yariv, J. D. O'Brien, P. D. Dapkus, I. Kim, Science, vol. **284**, 1819 (1999)

[6] J. Hwang, H. Ryu, O. Song, I. Han, H. Song, H. Park, Y. Lee, D. Jang, Appl. Phys. Lett. **76,** 2982, (2000)

[7] M. Loncar, T. Yoshie, A. Scherer, P. Gogna, Y. Qiu, Appl. Phys. Lett. **81**, 2680 (2003)

[8] T. Yoshie, M. Loncar, A. Scherer, Y. Qiu, Appl. Phys. Lett. **84**, 3543 (2004)

[9] H. Altug, J. Vučković, Appl. Phys. Lett. **84**, 161 (2004)

[10] M. Loncar, A. Scherer, Y. Qiu, Appl. Phys. Lett. **82**, 4648 (2003)

[11] S. Noda, M. Yokoyama, M. Imada, A. Chutinan, M. Mochizuki, Science **293**, 1123, (1999)

[12] H. Ryu, H. Park, Y. Lee, J. Selected Topics in Quantum Electronics **8**, 891 (2002)

**FIGURE CAPTIONS**

**Figure 1.** (a) Experimental setup used in measurement of the CPCRA band diagram. The explanation of symbols is as follows: λ/2 - half-wave plate, BS - beam splitter, PBS - polarizing beam splitter, OL - objective lens, IR-cam - infrared camera, and D - detector. Rotations of the structure around the y- and z-axes are controlled by two rotation stages. (b) SEM pictures of the fabricated CPCRA with $A=3a$.

**Figure 2.** (a) Measured band diagram (from the transmission spectrum) of the CPCRA from Fig. 1 for the coupled x-dipole mode in the ΓX direction. The dark blue stripe is the location of the coupled band. The horizontal axis is the tilt angle around the y-axis, which corresponds to different $k_x$-values; the vertical axis is the wavelength. The insets show the transmission spectrum at 0° tilt, i.e., at the Γ point, and the magnetic field ($B_z$) pattern of the coupled x-dipole mode at the X-point. (b) The measured band diagram for coupled y-dipole mode. The inset shows the magnetic field ($B_z$) pattern at the X point.

**Figure 3.** Comparison of the experiment with the FDTD simulations including structural astigmation and underlying oxide layer. Solid lines with filled markers are experimental results and dashed lines with empty markers are simulation results. The horizontal axis corresponds to the probed fraction of the indicated crystal direction and the vertical axis is the wavelength. (a) Dipole bands in the ΓX direction. Triangles and squares denote one set of x- and y- dipole bands (also shown in Fig. 2), while inverted triangles and circles denote the other set, obtained after rotating the structure by 90° around the z-axis; (b) x-like and y-like dipole bands in the ΓM direction; (c) Quadrupole band in the ΓX direction. The middle subfigure shows the magnetic field ($B_z$) pattern of the quadrupole mode at the X point. The subfigure on the right shows the unit cell and high symmetry directions.

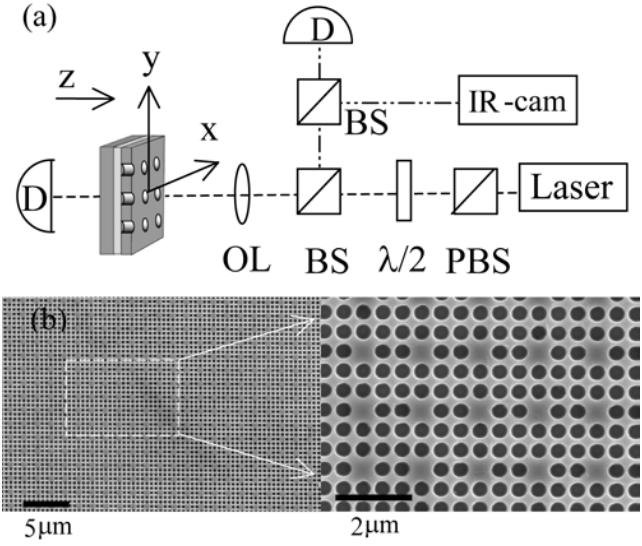

FIGURE 1

Authors: H. Altug, J. Vučković

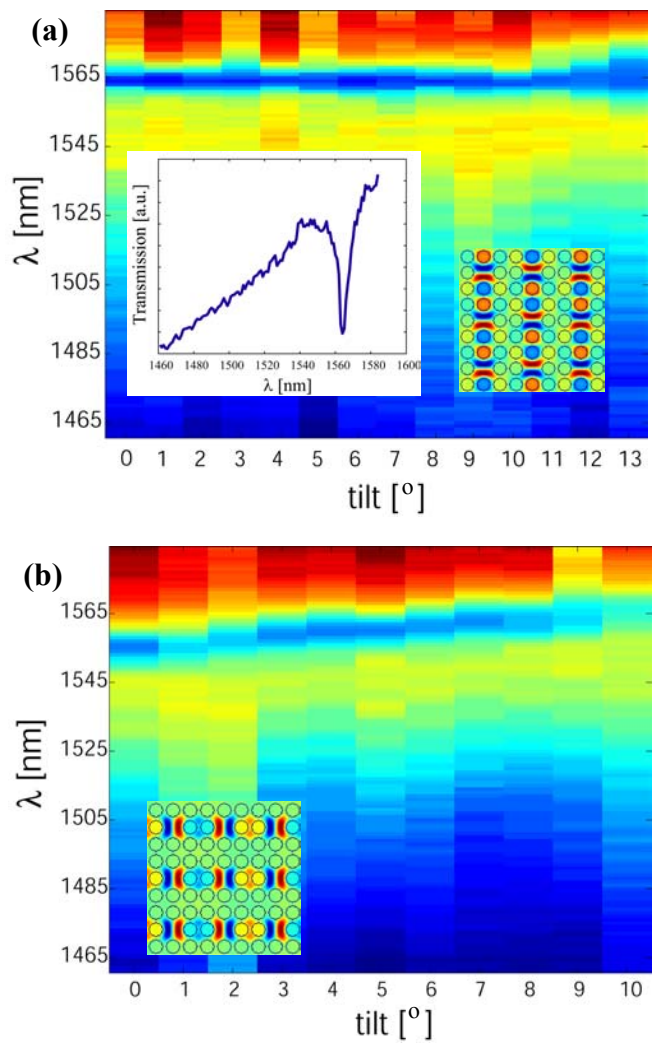

FIGURE 2

Authors: H. Altug, J. Vučković

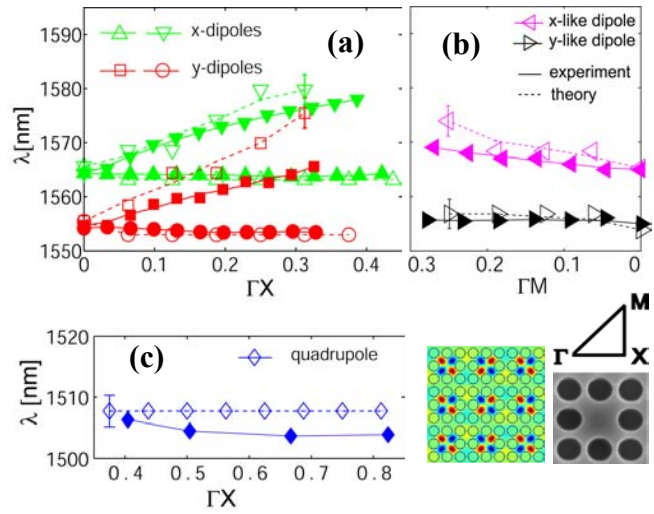

FIGURE 3

Authors: H. Altug, J. Vučković